\documentclass[]{aastex631}

\usepackage{natbib}

\received{23 April, 2022}
\revised{27 April, 2022}
\accepted{6 May, 2022}

\graphicspath{{./}{figures/}}

\begin{document}

\title{Spectroscopic Signature of a Re-established Accretion Disk in  Symbiotic -like Recurrent Nova RS Ophiuchi}
 
\correspondingauthor{S. N. Shore}
\email{steven.neil.shore@unipi.it}

\author[0000-0002-0786-7307]{Alessandra Azzollini}
\affiliation{ Lehrstuhl f\"ur Astronomie, 
 Universit\"at W\"urzburg\\ 
 Emil-Fischer-St. 31, W\"urzburg, 97074, Germany\\
and\\ Dipartimwnto  di Fisica. Universit\'a di Pisa \\
largo B. Pontecorvo 3 \\
Pisa 56127, Italy }

\author[0000-0003-1677-8004]{Steven N. Shore }
\affiliation{Dipartimento di Fisica,Uniersit\'a di Pisa and INFN - Sezione di Pisa \\
largo B. Pontecorvo 3 \\
Pisa 56127 Italy}

\author[0000-0003-4650-4186]{N. Paul Kuin }
\affiliation{ Mullard Space Science Laboratory\\     
 University College London\\
Holmbury St. Mary, Dorking, United Kingdom}

\begin{abstract}

A novel method is presented which can pin down the time the accretion disk re-established 
itself in the RS Oph system after it experienced a nova disruption. The method is based on 
the re-ionisation of the ejecta by photoionisation from the radiation released in the boundary 
layer from accretion.

\end{abstract}

 
\section{Introduction} \label{sec:intro}

 Among the recurrent novae, one small subgroup erupts within the wind of a close red giant companion (see, e.g., 
 \citet{Bode10}, \citet{anup13}, \citet{Darn20}).  These systems are very different than the classical, compact novae for which the ejecta are expanding freely once the explosion terminates.  In these symbiotic-like binaries, the accretion is, at least in part, driven by capture of the imbedding wind by the companion white dwarf.  A disk forms and the accretion proceeds similarly to the compact, cataclysmic type systems.  The main difference is in the post-eruption behavior of the expelled mass.  The ejecta plot into a dense, stratified environment in which a strong leading shock is formed that also modifies the ionization and density structure of the wind.  An important question, as yet unresolved, is how the system relaxes back to the pre-outburst accretion state and when the disk -- that disrupts during the explosion, reforms.  

\section{Disks and stochastic variability: flickering}

\citet{Zam220} and \citet{Rom22} have recently reported that flickering -- taken as the signature of vigorous accretion in cataclysmic binaries -- has reappeared in the post-outburst symbiotic-like recurrent nova RS Ophiuchi.  The current outburst has been followed across the spectrum.  The previous outburst, in 2006, was also extensively covered, especially well with the UVOT grism on the {\it Neil Gehrels Swift Observatory} from day 30 for about one year in the range 1700 - 7000 \AA with a resolution of about 150 \citep{Azz21}.  The 2021 outburst of RS Oph began on Aug 8 (MJD 59674), corresponding to orbital phase 0.$^p$72, with zero being inferior conjunction of the red giant \citep{Brandi09}.  The 2006 outburst began on Feb. 12, at orbital phase 0.$^p$26.  Thus, views through the wind of the giant of the white dwarf and its environs were completely different, in the 2006 event the system was seen without the intervening obscuration, around 0.$^p$65 and in this latest the orbital phase was about 0.$^p$11.  Thus, we concentrate on that spectra sequence. 

The spectrum was composite, consisting of contributions from the shock and associated ejecta from the nova, emission lines from the wind of the red giant produced by the UV precursor generated by the shock, and the additional ionization following the appearance of a soft X-ray source below 1 keV due to emission from the pointlike white dwarf.  The latter appeared only after the combined neutral column density of the ejecta and wind  was sufficiently reduced by the advance of the ionization front.  The He II 4686\AA\ and Balmer lines were especially well observed throughout the interval and we show the variations of H$\delta$ and He II in figure 1.    Overplotted we display the latest flux measurements from the 2021 eruption that follow precisely the same pattern.

\begin{figure}[ht!]
\plotone{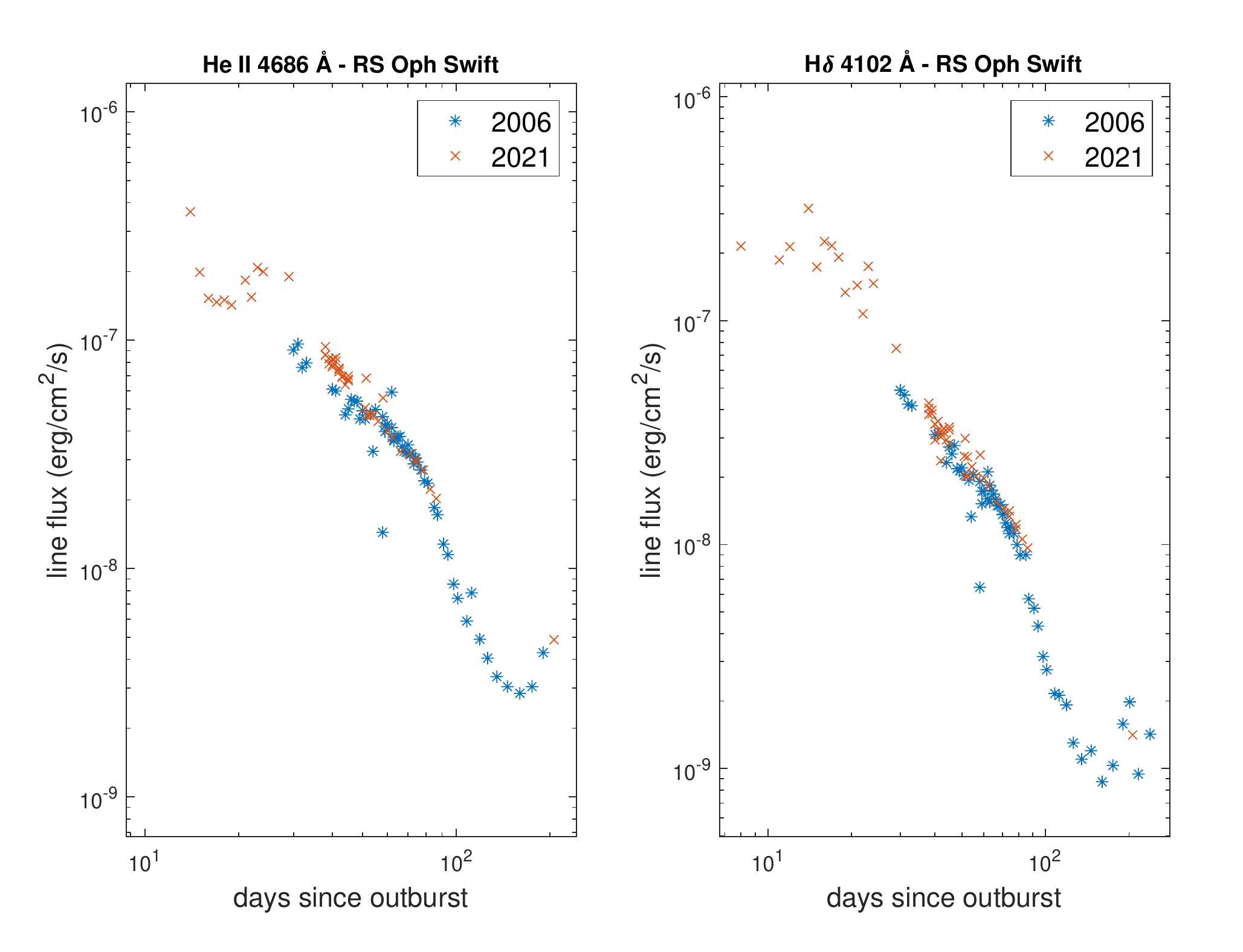}
\caption{{\it Swift} UVot grism development of H$\delta$ and He II 4686\AA\ emission lines in RS Oph comparing the 2006 and 2021 outbursts (uncorrected for extinction).  The scaling between the two events is a constant factor of 22.   See text for discussion.  \label{fig:general}}
\end{figure}

The variations can be phenomenologically separated into four developmental states.  The first is from the shock during its free expansion stage and its subsequent Sedov-Taylor expansion.  There was also an accumulation of mass from the environment, a peculiarity of these systems in contrast to classical nova explosions that expand ballistically from the start.  This snowplow increased the mass of the ejecta and slowed them rapidly.  The second stage, from about day $\sim 75$, 
is breakout, when the further mass loading of the ejecta is negligible but the gas is continually ionized by  the residual nuclear burning on the white dwarf (the so-called supersoft source).  The third stage, from around day 100,  is when following the turnoff of the central source, the ballistic expansion rate competes with the recombinations \citep{Sho96}.  The final, fourth stage is when the accretion disk has reformed and re-establishes the boundary layer and again photoionizes the ejecta.  
  
It is this last stage that is signalled by a renewed increase in the strength of the higher Balmer and ionized permitted lines, especially He II 4686\AA.  In the 2006 sequence, this last stage began with minimum strength for the emission lines at after around day 150 for the higher Balmer sequence and about day 160 for He II.  In the current outburst that corresponds to after about 2022 Feb. 15, when \citet{Mar22} did not detect optical flickering.  The formation of a disk is assured in these systems by the continued accretion from the companion's wind but the photometric flickering requires the formation of a steady state disk and boundary layer.  Therefore, we propose that the beginning stages of disk formation were consistent with the last stage of emission line decline, thus providing an independent means for establishing how and when  the white dwarf recommences its buildup to the next explosion.  The delay between the photometric and spectroscopic indicators can be attributed to the time required for viscosity to bring the accretion disk to a steady state. 
 
\begin{acknowledgments}
We thank Kim Page, Jordi Jos\'e,and Marco Bellomo for discussions.   NPMK acknowledges support by the UKSA.
.\end{acknowledgments} 
%

\vspace{5mm}
\facilities{Swift(XRT and UVOT), AAVSO}


\begin{thebibliography}{}
\bibitem[Azzollini (2021)]{Azz21} Azzollini, A. 2021, MSc thesis, Physics, Univ. di Pisa (https://etd.adm.unipi.it/t/etd-07012021-023228/)
\bibitem[Anupama (2013)]{anup13} Anupama, G. 2013, IAUS, 281, 154
\bibitem[Bode (2010)]{Bode10} Bode, M. 2010, AN, 331, 160
\bibitem[Brandi et al. (2009)]{Brandi09} Brandi, E., Quiroga, C, Mikolajewska, J, et al.  2009, A\&A, 497, 815
\bibitem[Darnley \& Henze.(1990)]{Darn20} Darnley, M. \& Henze, M. 2020, AdSpR, 66, 1147
\bibitem[Marchev et al. (2022)]{Mar22} Marchev, O. et al. 2022, ATel 15296
\bibitem[Romanov (2022)]{Rom22} Romanov, F. 2022, ATel 15338
\bibitem[Shore et al. (1996)]{Sho96} Shore, S. N., Sonneborn, G., 1996, ApJ, 463, L21
\bibitem[Zamanov et al. (2022)]{Zam220} Zamanov, R. et al. 2022, ATel 15330

\end{thebibliography}
\bibliographystyle{aasjournal}



\end{document}